\documentclass[sigconf]{acmart}

\usepackage[utf8]{inputenc}
\usepackage{booktabs,multirow}
\usepackage{enumitem}
\usepackage{color}
\usepackage{rotating}
\usepackage{amsmath}
\usepackage{array}
\usepackage[linesnumbered,algoruled,boxed,lined]{algorithm2e}
\usepackage{balance}
\newcommand{\ie}{{\it i.e.}}
\newcommand{\eg}{{\it e.g.}}

\newcommand{\ul}{\underline}{}

\AtBeginDocument{%
  }

\copyrightyear{2023}
\acmYear{2023}
\setcopyright{acmlicensed}\acmConference[SIGIR '23]{Proceedings of the 46th International ACM SIGIR Conference on Research and Development in Information Retrieval}{July 23--27, 2023}{Taipei, Taiwan}
\acmBooktitle{Proceedings of the 46th International ACM SIGIR Conference on Research and Development in Information Retrieval (SIGIR '23), July 23--27, 2023, Taipei, Taiwan}
\acmPrice{15.00}
\acmDOI{10.1145/3539618.3591966}
\acmISBN{978-1-4503-9408-6/23/07}

\begin{document}

\title{ConQueR: Contextualized Query Reduction using Search Logs}

\author{Hye-young Kim}\authornote{Equal contribution}
\affiliation{
  \institution{Sungkyunkwan University}
  \country{Republic of Korea}}
\email{khyaa3966@skku.edu}

\author{Minjin Choi}\authornotemark[1]
\affiliation{
  \institution{Sungkyunkwan University}
  \country{Republic of Korea}}
\email{zxcvxd@skku.edu}

\author{Sunkyung Lee}
\affiliation{
  \institution{Sungkyunkwan University}
  \country{Republic of Korea}}
\email{sk1027@skku.edu}

\author{Eunseong Choi}
\affiliation{
  \institution{Sungkyunkwan University}
  \country{Republic of Korea}}
\email{eunseong@skku.edu}

\author{Young-In Song}
\affiliation{
  \institution{NAVER Corp.}
  \country{Republic of Korea}}
\email{song.youngin@navercorp.com}

\author{Jongwuk Lee}\authornote{Corresponding author}
\affiliation{
  \institution{Sungkyunkwan University}
  \country{Republic of Korea}}
\email{jongwuklee@skku.edu}

\renewcommand{\shortauthors}{Hye-young Kim, et al.}

\begin{abstract}
Query reformulation is a key mechanism to alleviate the linguistic chasm of query in ad-hoc retrieval. Among various solutions, \emph{query reduction} effectively removes extraneous terms and specifies concise user intent from long queries. However, it is challenging to capture hidden and diverse user intent. This paper proposes \emph{\textbf{Con}textualized \textbf{Que}ry \textbf{R}eduction (ConQueR)} using a pre-trained language model (PLM). Specifically, it reduces verbose queries with two different views: \emph{core term extraction} and \emph{sub-query selection}. One extracts core terms from an original query at the \emph{term level}, and the other determines whether a sub-query is a suitable reduction for the original query at the \emph{sequence level}. Since they operate at different levels of granularity and complement each other, they are finally aggregated in an ensemble manner. We evaluate the reduction quality of ConQueR on real-world search logs collected from a commercial web search engine. It achieves up to 8.45\% gains in exact match scores over the best competing model.~\footnote{All the source codes are available at https://github.com/HyeYoung1218/ConQueR.}
\end{abstract}

\vspace{-3mm}


\vspace{-2mm}


\maketitle

\section{Introduction}\label{sec:introduction}

Query reformulation refers to the refinement of a seed query to obtain desired results by bridging the linguistic chasm of query~\cite{ooiMHS15SurveyExpansionSuggestionRefinement}. Empirically, approximately 28\% of queries are revised from the original query~\cite{PassCT06pictureofsearch}. It can be categorized into three approaches: (i) \emph{query reduction}~\cite{GuptaB15ReductionSurvey1, BenderskyC08ReductionSurvey2, BenderskyMC11ReductionSurvey3}, removing extraneous query terms, (ii) \emph{query expansion}~\cite{AzadD19ExpansionSurvey1, CarpinetoR12ExpansionSurvey2}, adding extra query terms, and (iii) \emph{query refinement}~\cite{VelezWSG97RefinementFastEffective, KoenemannB96RefinementInteraction, Anick03RefinementTerminological, SadikovMWH10RefinementClustering}, modifying the original query terms. Among them, query reduction is particularly beneficial for reducing long queries to better reflect user intent by narrowing down a search space. Interestingly, about 27\% of zero-hit queries, where users do not click on any documents, can be turned into successful ones by removing unnecessary query terms on an e-commerce site~\cite{AmemiyaMFS21zero-hit}. In this sense, query reduction is a simple yet effective way to reflect users' information needs.

Existing query reduction methods have been widely studied in two directions. First, conventional studies~\cite{ZukermanRW03ExpansionReductionICTAI, YangF17ShorterICTIR, XueHC10CRFCIKM, KumaranC09LongQueryPredictorSIGIR, MaxwellC13PseudoRelevanceSIGIR, KoopmanCZ17SIGIRGeneratingClinical, ChaaNB16Reductionbooksearch} focus on identifying salient features to improve the quality of search results in a limited resource, \eg, the TREC dataset~\cite{Clarke10TREC2010WebTrack}. Given a query, they define an optimal reduction by measuring the retrieval effectiveness for all reductions. Although they improve retrieval performance, it is difficult to measure the generalized performance due to the limited dataset (\eg, query-document relevance not being fully annotated). It is thus non-trivial to evaluate whether the defined optimal reduction effectively captures the actual user intent. Second, several studies~\cite{JonesF03SingleTermDeletionSIGIR, YangPSS14LargeSearchLogsECIR, HalderCCRWA20QueryReductionBiGRU} collect original and reduced query pairs from user search logs and analyze the characteristics of the dropped terms. Since they utilize real-world search logs, it is possible to capture hidden user intent in general scenarios. To fully exploit search logs, we need to consider three factors: (i) the meaning of query terms can vary depending on the context, (ii) the original and reduced queries are semantically consistent since they imply the same intent, and (iii) there may be inevitable noise in search logs. However, most existing methods are based on simple rule-based or RNN-based methods and do not take them into account well.

The pre-trained language model (PLM), \eg, BERT~\cite{DevlinCLT19BERT}, has recently achieved incredible performance improvements in the IR community. For document ranking, monoBERT~\cite{NogueiraC19monoBERT} adopts BERT to capture complex matching signals between a query and documents. For query expansion, CEQE~\cite{NaseriDYA21CEQEECIR} and BERT-QE~\cite{ZhengHHH0Y20BERTQE} use BERT to select terms or chunks in documents that are contextually similar to the query based on pseudo-relevance feedback. Inspired by these studies, we attempt to leverage the PLM to better capture the contextual meaning of queries.

To this end, we propose a novel PLM-based query reduction model, called \emph{\textbf{Con}textualized \textbf{Que}ry \textbf{R}eduction (ConQueR)}, using real-world search logs. We develop two methods with different views: \emph{core term extraction} and \emph{sub-query selection}. (i) For core term extraction, it takes the original query as input and distinguishes whether each term is important at the term level. That is, it validates whether a given term deviates from user intent in the context. (ii) For sub-query selection, it takes the original query and its candidate sub-query as input. It then determines whether a given sub-query sufficiently represents the original query at the sequence level. Hence, it evaluates whether the candidate sub-query is semantically complete and it is suitable for reflecting the original intent.

Finally, we aggregate the two methods because they complement each other by tackling query reduction at different levels. For example, since the core term extraction method determines importance at the term level, it may remove a subset of the terms that only make sense when they exist together (\eg, ``Silicon Valley''). Meanwhile, the sub-query selection method at the sequence level helps preserve the coherent semantics between the original query and its sub-query; however, it may also yield a high relevance score even if the sub-query is not sufficiently reduced. Therefore, it is beneficial to combine them to identify the most appropriate sub-query among the candidates. We additionally adopt a robust training strategy with a \emph{truncated loss} to deal with noisy search logs. Experimental results show that the proposed model outperforms existing models with gains of up to 8.45\% on real-world search logs.

\section{Proposed Model}

\begin{figure}
\includegraphics[width=0.85\linewidth]{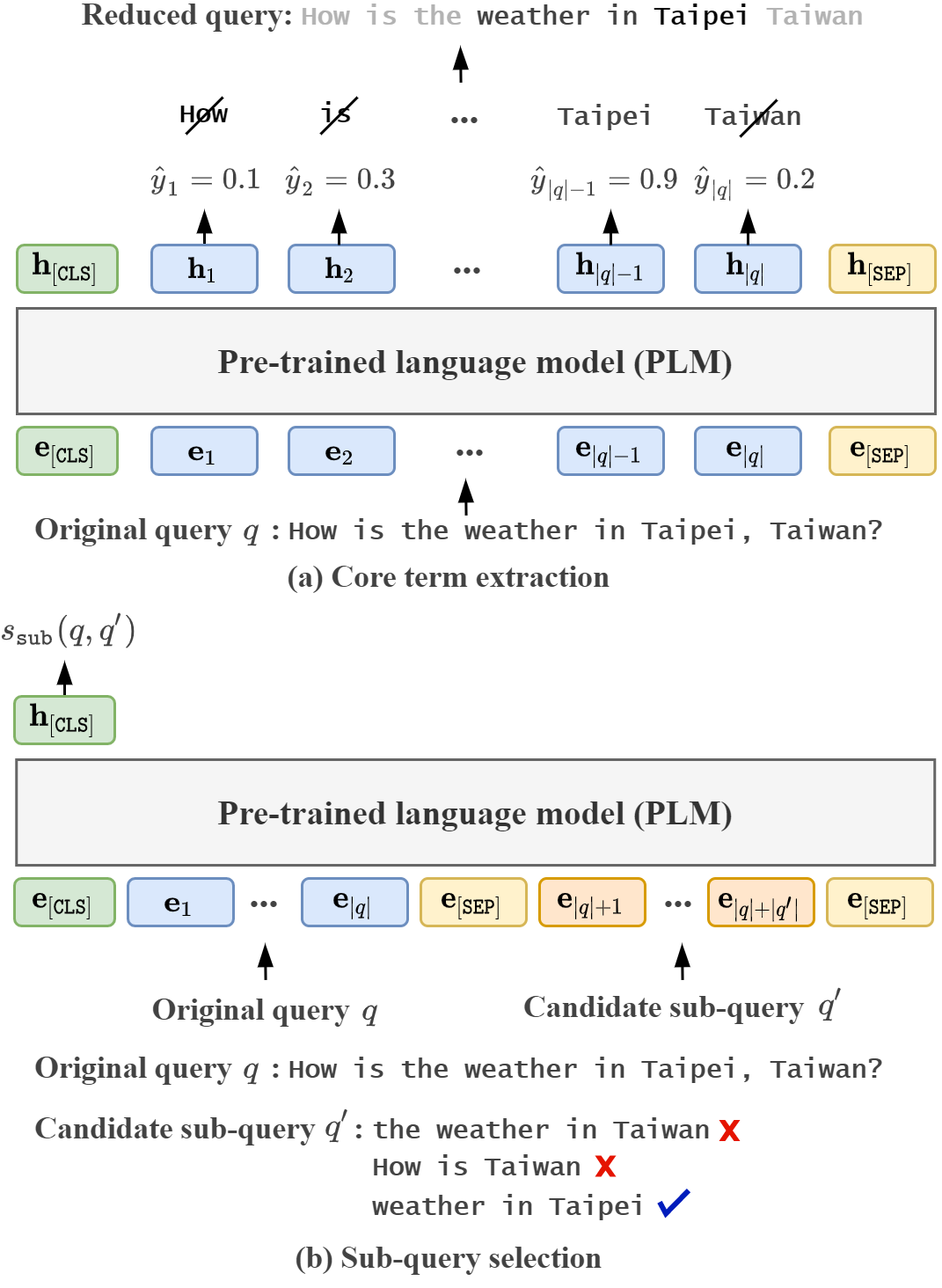}
\caption{Model architecture of ConQueR. Note that while only the original query $q$ is used as input for core term extraction, a pair of the original query $q$ and the candidate sub-query $q'$ is used for sub-query selection.}
\label{fig:architecture}
\vskip -0.1in
\end{figure}

In this section, we propose a novel query reduction model, namely \emph{\textbf{Con}textualized \textbf{Que}ry \textbf{R}eduction (ConQueR)}. To effectively reduce a long query while preserving the semantics of the original query, ConQueR exploits the contextualized representations of queries using PLMs~\cite{DevlinCLT19BERT, ClarkLLM20ELECTRA} with two different views. As shown in Figure~\ref{fig:architecture}, one extracts the core terms from the query at the term level; it determines whether each term is necessary and can leave only the appropriate terms that capture the user's intent. The other selects the most appropriate sub-query among the candidate sub-queries at the sequence level; it evaluates whether a given sub-query is suitably reduced or not by measuring the coherence of the original and the sub-query. It evaluates whether the sub-query properly captures the semantics of the user intent. Since they learn query reduction at different granularities, we finally combine them to produce synergies in performance. Furthermore, we adopt a robust training strategy with a \emph{truncated loss} to deal with noisy samples in search logs.

\subsection{Core Term Extraction}\label{sec:coreterm_architecture}
The core term extraction method determines the importance of each term in the query. It effectively predicts important terms and drops all nonessential terms.

\noindent\textbf{Model architecture.} Given a query $q$, we first tokenize it and take them as input, including two special tokens [\texttt{CLS}] and [\texttt{SEP}].
\begin{equation}
\left[ \mathbf{e}_{[\texttt{CLS}]}, \mathbf{e}_{1}, \dots, \mathbf{e}_{|q|}, \mathbf{e}_{[\texttt{SEP}]} \right],
\end{equation}
where each embedding vector $\mathbf{e}_i$ is combined with a positional embedding. Each embedding vector is passed into the transformer encoder and processed into hidden vector $\mathbf{h}_{i} \in \mathbb{R}^{k}$ for $i \in \{[\texttt{CLS}], 1,$ $\dots, |q|, [\texttt{SEP}]\}$ where $k$ is the dimension of embedding vectors. Note that we follow the conventional structure in BERT~\cite{DevlinCLT19BERT}.

We evaluate whether each query term is important or not at a term level. That is, each hidden vector $\mathbf{h}_{i}$ is projected into term importance score $\hat{y}_i$.
\begin{equation}
\label{eq:y_hat}
\hat{y}_i = \sigma(\mathbf{w}_{c} \mathbf{h}_{i} + b_{c}),
\end{equation}
where $\sigma(\cdot)$ is the sigmoid function, $\mathbf{w}_{c} \in \mathbb{R}^{1 \times k}$ is a weight parameter, and $b_{c}$ is a bias. Here, we determine whether each term should be retained or dropped; the retained terms indicate core terms reflecting the user's actual information need, and the dropped terms are unnecessary terms disrupting desired search results.

\noindent\textbf{Training \& inference.} As the core term extraction method learns reduction in a term level, we adopt a binary cross-entropy loss for each term.
\begin{equation}
\label{eq:coreterm_loss}
\mathcal{L}_{\text{core}} = - \sum_{i=1}^{|q|}  {y}_i\log\hat{y}_i + (1 - y_i) \log (1 - \hat{y}_i),
\end{equation}
\noindent
which is defined over an original and reduced query pair $(q, q^+) \in \mathcal{R}$. $\mathcal{R}$ is a training set of query pairs. $y_i$ is 1 if ground-truth reduced query $q^+$ contains $i$-th query term $q_i$, and otherwise 0. At inference, we remove the terms with an importance score $\hat{y}_i$ less than 0.5.

\subsection{Sub-query Selection}\label{sec:subquery_architecture}

The sub-query selection method takes the original query and a candidate sub-query as input to the transformer encoder. It determines whether the given sub-query is suitably reducing the original query. The coherence of two queries is evaluated by the cross-encoder mechanism, which performs all-to-all interactions across all terms from queries, as discussed in \cite{NogueiraC19monoBERT}.

\noindent\textbf{Model architecture.} Given the original query $q$ and a candidate sub-query $q'$, we tokenize and pass them into the transformer encoder. 
\begin{equation} 
\left[ \mathbf{e}_{[\texttt{CLS}]}, \mathbf{e}_{1}, \dots, \mathbf{e}_{|q|}, \mathbf{e}_{[\texttt{SEP}]}, \mathbf{e}_{|q|+1}, \dots, \mathbf{e}_{|q|+|q'|}, \mathbf{e}_{[\texttt{SEP}]} \right].
\end{equation}

By passing them to the PLM, we utilize contextualized hidden vector $\mathbf{h}_{[\texttt{CLS}]}$ to quantify the coherence of two sequences. It is projected into a sub-query score $s_{\text{sub}}(q,q')$ for the pair of the query $q$ and its sub-query $q'$.
\begin{equation}
\label{eq:s_sub}
s_{\text{sub}}(q,q') = \mathbf{w}_{s} \mathbf{h}_{[\texttt{CLS}]} + b_{s},
\end{equation}
where $\mathbf{w}_{s} \in \mathbb{R}^{1 \times k}$ is a weight parameter, and $b_{s}$ is a bias.

\noindent\textbf{Training \& inference.} 
As the sub-query selection method learns reduction at the sequence level, we adopt a negative log-likelihood of the positive query pair ($q, q^+$).
\begin{equation}
\label{eq:subquery_loss}
\mathcal{L}_{\text{sub}} = - \log \frac{\exp(s_{\text{sub}}(q,q^+))}{\exp(s_{\text{sub}}(q,q^+)) + \sum_{ q^- \in \mathcal{N}(q)} \exp(s_{\text{sub}}(q,q^-))},
\end{equation}
where $\mathcal{N}(q)$ is a set of the negative sub-queries for the original query $q$. Although it is possible to utilize all sub-queries except for the ground truth as the negative set, it significantly increases the training time. We thus sample a subset of them. (Empirically, we set the size of $\mathcal{N}(q)$ to five and re-sample them for each epoch.)

We infer the best sub-query with a greedy search instead of enumerating all the sub-queries, \ie, all subsets of query terms, which induces huge complexity. We first generate all possible candidates that delete only a single term from the original query and compute the score for these $|q|$ candidates to find the top-1 sub-query. Likewise, we generate new $|q|-1$ candidates by deleting one word from the previous top-1 and scoring them. We repeatedly generate candidates and select the top-1 sub-query until the top-1 sub-query is not changed.

\subsection{Aggregating Two Methods} \label{sec:Aggregation}

Since the two methods capture contextual information at multifaceted levels, \ie, term and sequence levels, they complement each other. However, it is non-trivial to aggregate them because they derive their reductions in distinct ways; one removes terms in the original query using term importance scores, while the other selects the most appropriate sub-query among the candidates using sub-query scores. To bridge the gap between the two methods, we additionally obtain sub-query scores from the core term extraction method. Specifically, we define a sub-query score using probabilities for terms that are present or deleted in the subquery $q'$. First, we derive the probability that each query term is retained (or removed) as follows:
\begin{equation}
\label{eq:P_core}
    p_{\text{core}}(q_i|q,q')=
    \begin{cases}
        \hat{y}_i, & \text{if} \ q_i \in q', \\
        1 - \hat{y}_i, & \text{otherwise},
    \end{cases}
    \ \text{for} \ i \in \{1, \dots, |q|\}.
\end{equation}
The term importance score $\hat{y}_i$ from ~\eqref{eq:y_hat} equals the retention probability, so that $1-\hat{y}_i$ means the removal probability. Then, we estimate the score of the sub-query $q'$ using by averaging the term probabilities.
\begin{equation}
\label{eq:s_core}
    s_{\text{core}}(q,q')=\frac{1}{|q|}\sum_{i=1}^{|q|}p_{\text{core}}(q_i|q,q').
\end{equation}

Finally, we aggregate the two methods by summing the scores, $s_{\text{sub}}(q,q')$ in Eq.~\eqref{eq:s_sub} and $s_{\text{core}}(q,q')$ in Eq.~\eqref{eq:s_core}.
\begin{equation}
\label{eq:s_final}
    s(q,q') = s_{\text{sub}}(q,q') + \alpha \cdot s_{\text{core}}(q,q'),
\end{equation}
where $\alpha$ is a hyperparameter to control the weight of $s_{\text{core}}(q,q')$. (When $\alpha$ = 4, we empirically observe the best performance.) To select the final sub-query, we use the same greedy search as in the sub-query selection method, \ie, we repeatedly create candidate sub-queries and select the top-1 sub-query until it is no longer changed.

\vspace{1mm}
\noindent
\textbf{Denoising training strategy}. For robust training, we adopt a loss-based denoising strategy for two methods. The \emph{truncated loss} is a technique for dynamically pruning large-loss samples during the training process~\cite{WangF0NC21TruncatedLossWSDM}. Assuming that the large-loss samples are noisy, we dismiss the top $\epsilon(T)$\% large-loss samples from the training set $\mathcal{R}$, where $\epsilon(T)$ is a drop rate function with respect to the training epoch $T$. Specifically, we define the drop rate function as $\epsilon(T)=min(\frac{{\epsilon_{max}}^{\gamma}}{\epsilon_N-1}(T-1), \epsilon_{max})$, where $\epsilon_{max}$, $\gamma$ and $\epsilon_{N}$ are hyperparameters that control the drop rate per epoch.

\section{Evaluation}\label{sec:Evaluation}

\subsection{Experimental Setup}\label{sec:setup}

\textbf{Datasets}. We collect the search logs from a commercial web search engine\footnote{NAVER Corp., Republic of Korea.}. The dataset consists of Korean query pairs, where two queries are successive in a session, and the latter is always a terminological subset of the former. The total number of query pairs is 239,976, while the number of unique original queries is 104,002. This means that each query is reduced to 2.3 different forms on average, reflecting the various query reformulations of users. We split them into three subsets, \ie, a training set (83,202, 80\%), a validation set (10,400, 10\%), and a test set (10,400, 10\%), depending on the number of unique original queries. To remove the noise of the validation and test sets, we use the query pairs where the original query is always reduced to the same sub-query and appears in two or more sessions.

\vspace{1.0mm} \noindent
\textbf{Competing models}. We adopt seven existing methods as baselines. Rule-based methods~\cite{YangPSS14LargeSearchLogsECIR, JonesF03SingleTermDeletionSIGIR} are (i) LEFTMOST (LM): deleting $N_q$ leftmost terms, (ii) RIGHTMOST (RM): deleting $N_q$ rightmost terms, (iii) DF: deleting the most frequently deleted $N_q$ terms on the training set, and (iv) CDF: deleting $N_q$ terms with the highest ratio of \#deletions/\#appearances on the training set. For DF and CDF, backoff schemes are noted after the semi-colon. For neural methods~\cite{HalderCCRWA20QueryReductionBiGRU, CaoCBTC21SEQUER}, GRU~\cite{HalderCCRWA20QueryReductionBiGRU} predicts the removal probability of each query term using a bi-directional GRU~\cite{Chung14GRU}. SEQUER~\cite{CaoCBTC21SEQUER} is a transformer~\cite{VaswaniSPUJGKP17Transformer} model that takes the original query as input and generates a reformulated query, and it is trained from scratch in the original paper setting. For a fair comparison, we initialize it with a PLM (BART$_{\text{base}}$~\cite{LewisLGGMLSZ20BART}) and denote it as SEQUER$_{\text{BART}}$. As an additional baseline, SEQUER$_{\text{GPT}}$ is a transformer decoder model initialized with GPT-2 (125M)~\cite{radford2019GPT2}. ConQueR$_{\text{core}}$, ConQueR$_{\text{sub}}$, and ConQueR$_{\text{agg}}$ indicate the core term extraction, the sub-query selection, and the aggregated model, respectively. Some methods~\cite{ZukermanRW03ExpansionReductionICTAI, YangF17ShorterICTIR, XueHC10CRFCIKM, KumaranC09LongQueryPredictorSIGIR, MaxwellC13PseudoRelevanceSIGIR} are excluded since the actual ranking of sub-queries, and various query features cannot be obtained from the search logs.

\vspace{1mm}
\noindent
\textbf{Evaluation metrics}. We use an exact match score (EM), accuracy (Acc), precision (P), recall (R), and F1-score (F1). For each query, EM is 1 if the reduction is correct and 0 otherwise. For Acc, we divide the number of correct terms by the number of terms in a query. To compute P, R, and F1, we set the retention as a positive class and the removal as a negative class. All evaluation metrics are computed for each query and averaged.

\vspace{1mm}
\noindent
\textbf{Reproducibility}. For the proposed models, we initialized them with ELECTRA$_{\text{base}}$~\cite{ClarkLLM20ELECTRA}. We used the Adam optimizer and set the maximum sequence length to 60 and 120 for ConQueR$_{\text{core}}$ and ConQueR$_{\text{sub}}$, respectively. We set the dropout ratio to 0.2 and the maximum epoch to 5. On the validation set, we ran a grid search for batch size, learning rate, and warm-up ratio for a linear scheduler, and set them to 32, 1e-5, and 0.2, respectively. For truncated loss, $\epsilon_{max}$, $\epsilon_{N}$, and $\gamma$ were set to 0.3, 4, and 2, respectively. For aggregation, $\alpha$ is tuned in \{$\frac{1}{8}$, $\frac{1}{4}$, $\frac{1}{2}$, 1, 2, 4, 8\}. Experimental results are averaged over five runs with different seeds. For rule-based methods, we set the number of reduced terms $N_q$ to 1 because the majority of queries exclude only one word.

\begin{table}[t]\small
\caption{Comparison of existing query reduction models and ours. The best and second-best models are marked in \textbf{bold} and {\underline{underlined}}. Significant differences ($p < 0.01$) between baselines and ConQueR$_{\text{agg}}$ are denoted with *.}
\label{tab:effectivness}
\begin{center}
\vspace{-3mm}
\begin{tabular}{c|ccccc}
\toprule
Models & EM & Acc & P & R & F1 \\ \midrule
LEFTMOST & 0.170* & 0.338* & 0.396* & 0.415* & 0.403* \\
RIGHTMOST & 0.693* & 0.794* & 0.801* & 0.838* & 0.814* \\
DF;RM & 0.596* & 0.738* & 0.761* & 0.801* & 0.775* \\
CDF;RM & 0.697* & 0.814* & 0.822* & 0.864* & 0.836* \\ \midrule
GRU & 0.752* & 0.882* & 0.885* & 0.920* & 0.893* \\
SEQUER$_{\text{BART}}$ & 0.833* & 0.884* & 0.894* & 0.901* & 0.894* \\
SEQUER$_\text{GPT}$ & 0.840* & 0.885* & 0.895* & 0.901* & 0.896* \\ \midrule
ConQueR$_\text{core}$ & 0.892 & 0.928 & \ul{0.935} & 0.934 & 0.932 \\
ConQueR$_\text{sub}$ & \ul{0.905} & \ul{0.929} & \ul{0.935} & \ul{0.939} & \ul{0.935} \\
ConQueR$_\text{agg}$ & \textbf{0.911} & \textbf{0.934} & \textbf{0.939} & \textbf{0.943} & \textbf{0.940} \\ \bottomrule
\end{tabular}
\end{center}
\vspace{-5mm}
\end{table}


\vspace{-1mm}
\subsection{Results and Analysis}\label{sec:results}

Table~\ref{tab:effectivness} shows the overall query reduction accuracy. The key observations are as follows: (i) All proposed models consistently outperform the baselines. Although SEQUER$_{\text{BART}}$ and SEQUER$_{\text{GPT}}$ are the most competitive baselines, they are less effective than ours in capturing contextual information. ConQueR$_{\text{agg}}$ surpasses the best competing model by 8.45\% gain in EM and 5.54\% gain in Acc. (ii) ConQueR$_{\text{agg}}$ shows higher accuracy than ConQueR$_{\text{core}}$ and ConQueR$_{\text{sub}}$, proving that the different characteristics of the two methods are fully exploited by aggregation. (iii) Simple rule-based methods show relatively lower accuracy than neural methods, suggesting the limitations of naive approaches that cannot account for the contextual meaning of query terms. If we set the number of removing terms $N_q$=2, RIGHTMOST and CDF;RM show 0.080 and 0.061 for EM, and if $N_q$ is greater than 2, the performance is severely degraded.

\begin{figure}
\centering
\begin{tabular}{c}
\includegraphics[width=0.44\textwidth]{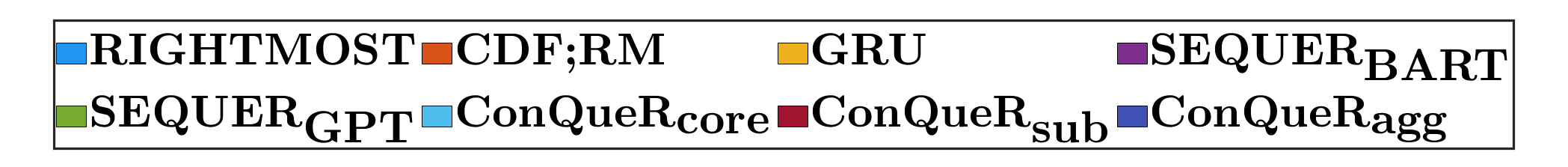}

\end{tabular}
\includegraphics[width=0.94\linewidth]{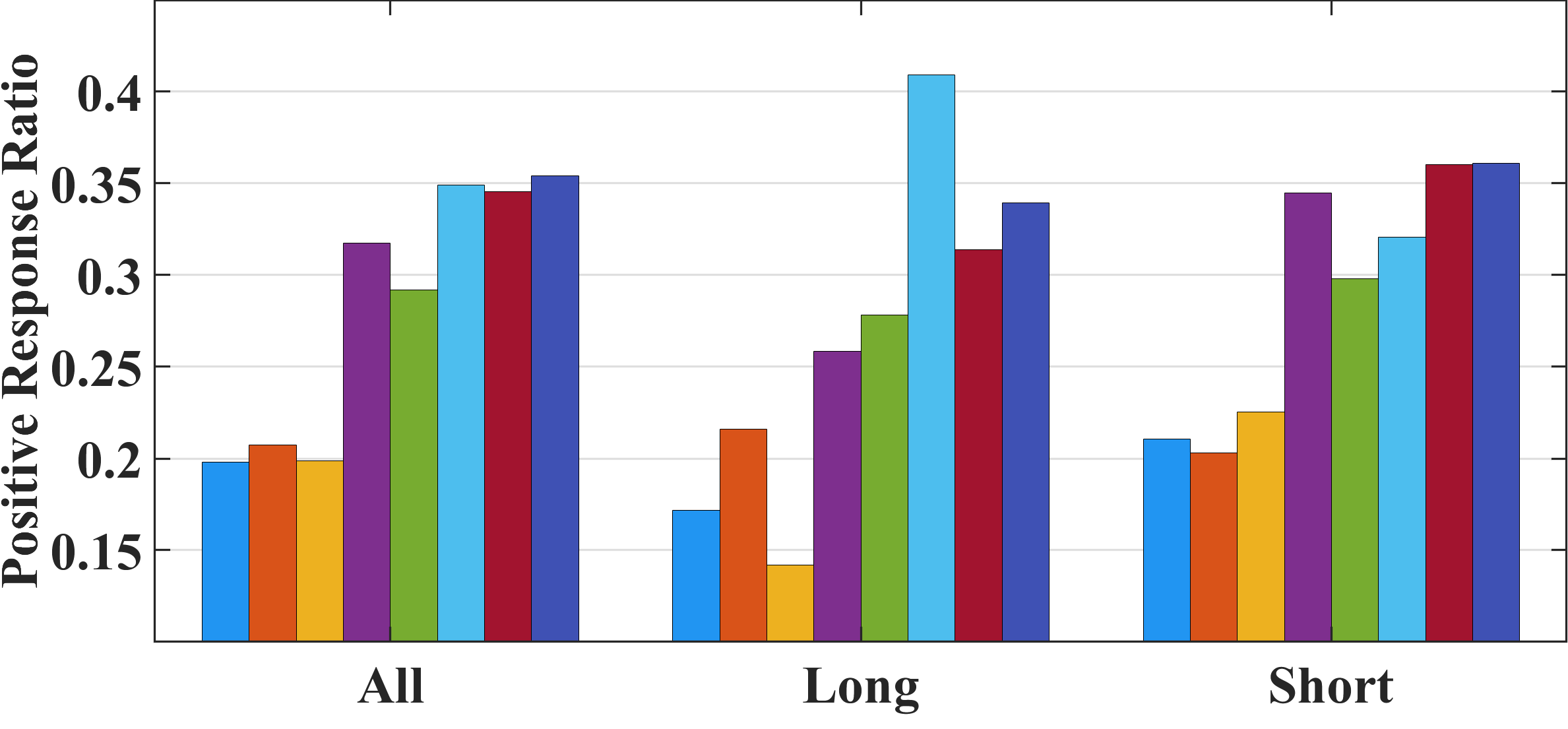}
\vspace{-3mm}
\caption{User study of existing query reduction models and ours. We reported user responses on 25 queries from 88 users. $\textit{Short}$ ($<$ 5) and $\textit{Long}$ ($\ge$ 5) consist of 13 and 12 queries, and are set based on the length of the original queries.}\label{fig:user_study}
\vskip -0.15in
\end{figure}

Lastly, we conduct a qualitative evaluation for query reduction. Given the original query and the reduced queries of anonymized models, users were asked whether each reduction was appropriate or not. We calculated the Positive Response Ratio as $\frac{{\sum_{u}{select_u}}}{\#users}$ and averaged it over all queries, where $select_u \in \{0,1\}$ indicates whether the user $u$ responded that the reduction is appropriate or not. As in Figure~\ref{fig:user_study}, ConQueR$_{\text{agg}}$ shows the best performance, indicating that the two methods are well aggregated. On average, about 35\% of the users think that ConQueR$_{\text{agg}}$ correctly reduces the original queries. Interestingly, ConQueR$_{\text{core}}$ and ConQueR$_{\text{sub}}$ perform better on both subsets of $\textit{Long}$ and $\textit{Short}$. This is because the longer the query, the more it needs to be reduced, and ConQueR$_{\text{core}}$ is more suited for deleting multiple terms. While ConQueR$_{\text{sub}}$ tends to give a high score to sub-queries that remove only a single term and are contextually similar to the original user intent.

\section{Conclusion}\label{sec:conclusion}

In this paper, we propose a novel query reduction model, namely \emph{\textbf{Con}textualized \textbf{Que}ry \textbf{R}eduction (ConQueR)}, with two different views. The core term extraction method distinguishes whether each term is essential at the term level. The sub-query selection method determines whether a sub-query is semantically coherent with the original query at the sequence level. To effectively consider query reduction in a multifaceted way, two methods are aggregated and complement each other. In addition, the truncated loss is employed to reduce the effect of noisy samples from search logs. Extensive results show that the proposed model outperforms existing models on the real-world search log dataset collected from a commercial web search engine. The user study verifies that it reduces queries more appropriately than the other methods.

\section*{Acknowledgement}
This work was supported by Institute of Information \& communications Technology Planning \& Evaluation (IITP) grant funded by the Korea government (MSIT) (No.2022-0-00680, 2022-0-01045, 2019-0-00421, 2021-0-02068, and IITP-2023-2020-0-01821).

\bibliographystyle{ACM-Reference-Format}
\balance
\bibliography{references}
\newpage
\appendix

\section{Additional Results}\label{sec:app_results}

\begin{table}[t!]\footnotesize
\caption{Comparison of query reduction models and ours according to the number of reduced terms. The best and second-best models are marked in \textbf{bold} and {\underline{underlined}}. The number in parentheses indicates the number of queries in each set. Significant differences ($p < 0.01$) between the baselines and ConQueR$_{\text{agg}}$ are denoted with *.}
\vspace{-3mm}
\label{tab:reduced_term}
\begin{tabular}{c|c|ccccc}
\toprule
Reduction type & Models & EM & Acc & P & R & F1 \\ \midrule
\multirow{11}{*}{\begin{tabular}[c]{@{}c@{}}Single term\\ deletion\\ (9,235)\end{tabular}} 
 & LEFTMOST & 0.191* & 0.344* & 0.407* & 0.407* & 0.407* \\
 & RIGHTMOST & 0.780* & 0.819* & 0.836* & 0.836* & 0.836* \\
 & DF;RM & 0.670* & 0.755* & 0.790* & 0.790* & 0.790* \\
 & CDF;RM & 0.784* & 0.834* & 0.854* & 0.854* & 0.854* \\ \cmidrule{2-7} 
 & GRU & 0.772* & 0.891* & 0.899* & 0.920* & 0.902* \\
 & SEQUER$_\text{BART}$ & 0.859* & 0.893* & 0.908* & 0.902* & 0.904* \\
 & SEQUER$_\text{GPT}$ & 0.864* & 0.896* & 0.909* & 0.905* & 0.906* \\ \cmidrule{2-7} 
 & ConQueR$_\text{core}$ & 0.909 & 0.937 & 0.946 & 0.939 & 0.941 \\
 & ConQueR$_\text{sub}$ & \ul{0.925} & \ul{0.942} & \ul{0.950} & \ul{0.947} & \ul{0.948} \\
 & ConQueR$_\text{agg}$ & \textbf{0.931} & \textbf{0.946} & \textbf{0.953} & \textbf{0.950} & \textbf{0.951} \\ \midrule
\multirow{11}{*}{\begin{tabular}[c]{@{}c@{}}Multi term\\ deletion\\ (1,165)\end{tabular}} 
 & LEFTMOST & 0.147* & 0.347* & 0.262* & 0.279* & 0.266* \\
 & RIGHTMOST & 0.714* & 0.819* & 0.775* & 0.802* & 0.783* \\
 & DF;RM & 0.542* & 0.756* & 0.721* & 0.754* & 0.730* \\
 & CDF;RM & 0.682* & 0.838 & 0.803* & 0.841* & 0.814* \\ \cmidrule{2-7} 
 & GRU & 0.588* & 0.810* & 0.778* & \textbf{0.925} & 0.823* \\
 & SEQUER$_\text{BART}$ & 0.634* & 0.811* & 0.784* & 0.892 & 0.821* \\
 & SEQUER$_\text{GPT}$ & 0.652* & 0.804* & 0.783* & 0.873* & 0.814* \\ \cmidrule{2-7} 
 & ConQueR$_\text{core}$ & \textbf{0.758} & \textbf{0.859} & \textbf{0.845} & \ul{0.896} & \textbf{0.861} \\
 & ConQueR$_\text{sub}$ & 0.743 & 0.826 & 0.815 & 0.873 & 0.833 \\
 & ConQueR$_\text{agg}$ & \ul{0.752} & \ul{0.840} & \ul{0.827} & 0.889 & \ul{0.847} \\ \bottomrule
\end{tabular}
\vspace{-2.5mm}
\end{table}

\begin{figure}
\includegraphics[width=0.9\linewidth]{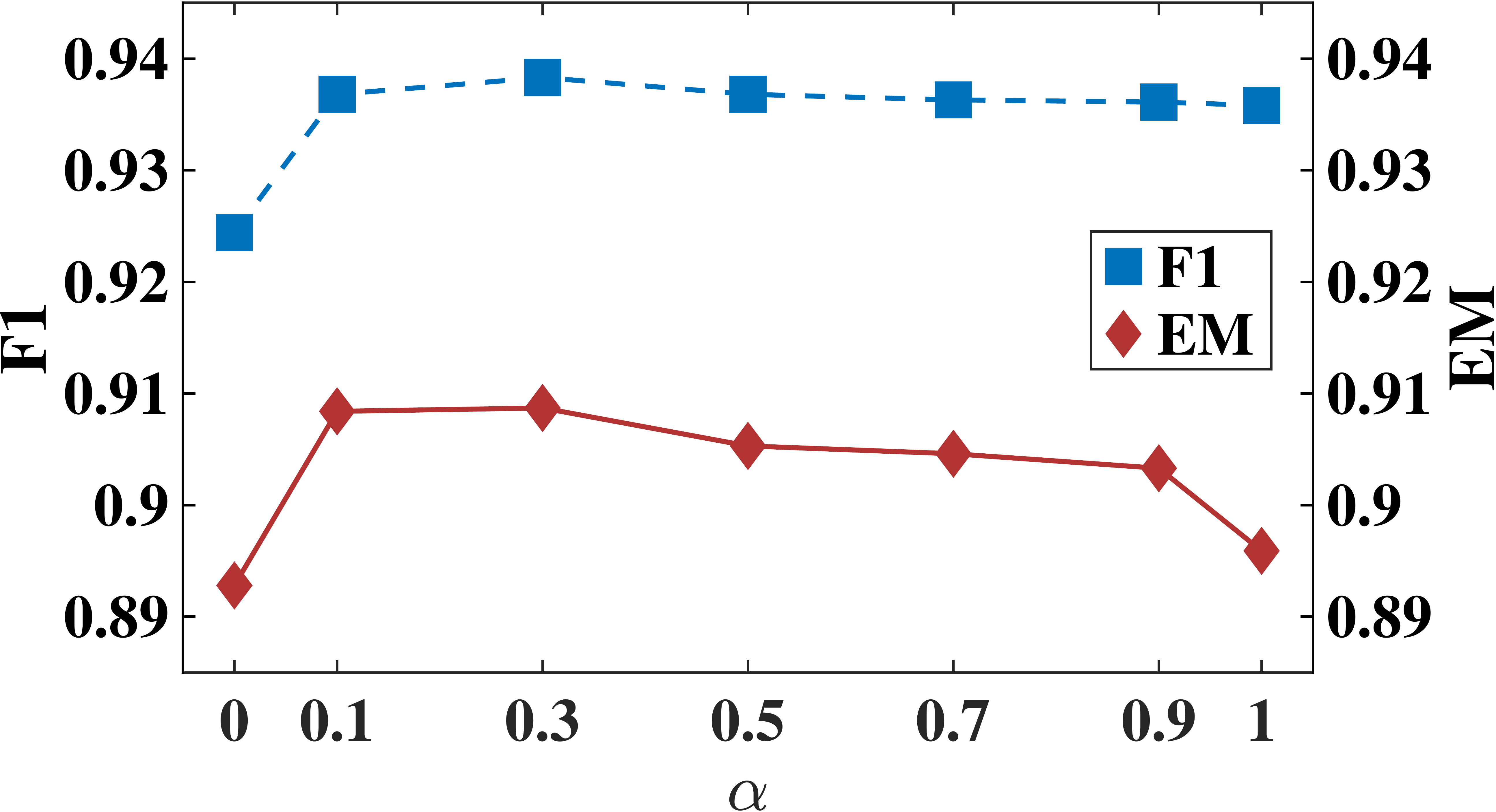}
\caption{Accuracy of ConQueR$_{\text{agg}}$ over varying $\alpha$.}
\label{fig:ensemble}
\end{figure}


\textbf{Effect of the number of reduced terms.} We further evaluate the proposed models depending on the number of reduced terms in Table~\ref{tab:reduced_term}. For rule-based methods, we set $N_q$=1 for single-term deletion and $N_q$=2 for multi-term deletion. ConQueR$_{\text{agg}}$ shows the best performance with gains of 7.75\% and 10.26\% in EM over the best baseline for single and multi-term deletion, respectively. ConQueR$_{\text{sub}}$ performs better for single term deletion, while ConQueR$_{\text{core}}$ performs better for multi-term deletion. This suggests that ConQueR$_{\text{sub}}$ tends to give a high score to sub-queries with only a single term removed because they are contextually similar to the original user intent.
The main observations are as follows: (i) The proposed models consistently achieve the best performance in EM for both single and multi-term deletion. 

\noindent
\textbf{Effect of the aggregating parameter.} Fig~\ref{fig:ensemble} shows the effect of hyperparameter $\alpha$ balancing the core-term component and sub-query component. When $\alpha$ is 0 or 1, ConQueR is equal to ConQueR$_{\text{CE}}$ or ConQueR$_{\text{SS}}$, respectively. F1 shows the highest accuracy when $\alpha$=1 and the highest accuracy in EM when $\alpha$ is about 0.9. Additionally, to validate the gain from the ensemble method, we also aggregate the two same components with different parameter initialization. Using two core-term or sub-query selection components achieve 0.896 and 0.893 in EM, respectively.


\end{document}